%% file: main.tex
\PassOptionsToPackage{capitalise}{cleveref}
\PassOptionsToPackage{table,xcdraw}{xcolor}

\documentclass[sigconf]{acmart}

\usepackage{colortbl}
\usepackage[utf8]{inputenc}
\usepackage[english]{babel}

\usepackage{graphicx}
\graphicspath{{./gfx/}}
\DeclareGraphicsExtensions{.pdf,.jpeg,.png,.eps}

\usepackage{tikz}
\usetikzlibrary{arrows,positioning,automata}
\usepackage{tikzscale}

\usepackage{paralist}

\usepackage{ifthen}
\newboolean{extended}
\setboolean{extended}{false}
\newcommand{\extended}[2]{\ifthenelse{\boolean{extended}}{#1}{#2}}

\usepackage{xcolor}

  \usepackage[caption=false,font=footnotesize]{subfig}
\usepackage[acronym]{glossaries}
\makeglossaries
\input{glossary}

\usepackage{csquotes}

\usepackage{tabulary}
\usepackage{multirow}
\usepackage{booktabs}

\newcommand{\para}[1]{\textbf{#1\@.}}

\renewcommand{\subsubsection}[1]{\para{#1}}

\copyrightyear{2017} 
\acmYear{2017} 
\setcopyright{acmcopyright}
\acmConference{MSWiM '17}{November 21--25, 2017}{Miami, FL, USA}
\acmPrice{15.00}
\acmDOI{10.1145/3127540.3127542}
\acmISBN{978-1-4503-5162-1/17/11}

\begin{document}

\title[Reverse Engineering Human Mobility in Large-scale Natural Disasters]{Reverse Engineering Human Mobility\\ in Large-scale Natural Disasters}

\newcommand{\SEEMOOaffiliation}{%
\affiliation{%
	\department{Secure Mobile Networking Lab}
	\institution{Technische Universität Darmstadt}
	\city{Darmstadt}
	\country{Germany}
}}

%
%
%

\author{Milan Stute, Max Maass, Tom Schons, and Matthias Hollick}
\SEEMOOaffiliation{}
\email{firstname.lastname@seemoo.tu-darmstadt.de}

\begin{abstract}
\glspl{DTN} have been around for more than a decade and have especially been proposed to be used in scenarios where communication infrastructure is unavailable. In such scenarios, \glspl{DTN} can offer a best-effort communication service by exploiting user mobility.
Natural disasters are an important application scenario for \glspl{DTN} when the cellular network is destroyed by natural forces.
To assess the performance of such networks before deployment, we require appropriate knowledge of human mobility.

In this paper, we address this problem by designing, implementing, and evaluating a novel mobility model for large-scale natural disasters.
Due to the lack of GPS traces, we reverse-engineer human mobility of past natural disasters (focusing on 2010 Haiti earthquake and 2013 Typhoon Haiyan) by leveraging knowledge of 126 experts from 71 \glspl{DRO}.
By means of simulation-based experiments, we compare and contrast our mobility model to other well-known models, and evaluate their impact on \gls{DTN} performance.
Finally, we make our source code available to the public.
\end{abstract}

%


\maketitle

\input{introduction}

\input{background}

\input{methodology}

\input{design}

\input{implementation}

\input{evaluation}

\input{conclusion}

\begin{acks}
	The authors would like to express their sincere gratitude towards all cooperating individuals from the following national and international organizations: %
		Caritas (AT),
		Berufsfeuerwehr Bochum (DE),
		Johanniter (DE),
		Red Cross (DE),
		Technisches Hilfswerk (DE),
		emergency.lu (LU),
		Humanitarian Intervention Team (LU),
		Ministry of Foreign Affairs (LU),
		Concern Worldwide,
		Plan International,
		ZOA International,
		Emer\-gency Telecommunication Cluster (ETC),
		Global Disaster Alert and Coordination System (GDACS),
		Office for the Coordination of Humanitarian Affairs (OCHA),
		United Nations Organizations (UNO),
		United Nations Office for Disaster Risk Reduction (UNISDR),
		World Food Program (WFP),
		and many others who wished to remain anonymous.
		
		This work has been co-funded by the LOEWE initiative within the NICER project, and by the DFG as part of project C.1 within the RTG 2050 ``Privacy and Trust for Mobile Users.''
\end{acks}

\bibliographystyle{ACM-Reference-Format}
\bibliography{library}

\end{document}

%% file: glossary.tex
\newacronym{DTN}{DTN}{Delay/Disruption-Tolerant Network}
\newacronym{ONE}{ONE}{Opportunistic Network Simulator}
\newacronym{EU}{EU}{European Union}
\newacronym{IFRC}{IFRC}{International Federation of Red Cross and Red Crescent Societies}
\newacronym{WKT}{WKT}{Well Known Text}
\newacronym{OSM2WKT}{OSM2WKT}{OpenStreetMap to WKT Conversion}
\newacronym{AODV}{AODV}{Ad hoc On-Demand Distance Vector Routing}
\newacronym{DSR}{DSR}{Dynamic Source Routing Protocol}
\newacronym{IPN}{IPN}{Inter Planetary Net}
\newacronym{BGAN}{BGAN}{Broadband Global Area Network}
\newacronym{GPS}{GPS}{Global Positioning System}
\newacronym{USD}{USD}{United States Dollar}
\newacronym{DWB}{DWB}{Doctors Without Borders}
\newacronym{GDP}{GDP}{Gross Domestic Product}
\newacronym{FEMA}{FEMA}{Federal Emergency Management Agency}
\newacronym{EMV}{EMV}{Emergency Management Victoria}
\newacronym{ERU}{ERU}{Emergency Response Unit}
\newacronym{ICT}{ICT}{Information and Communications Technology}
\newacronym{NASA}{NASA}{National Aeronautics and Space Administration}
\newacronym{RTT}{RTT}{Round Trip Time}
\newacronym{PRoPHET}{PRoPHET}{Probabilistic Routing Protocol for Intermittently Connected Networks}
\newacronym{SRO}{SRO}{Search and Rescue Operation}
\newacronym{USRT}{USRT}{Urban Search and Rescue Team}
\newacronym{USAR}{USAR}{Urban Search and Rescue}
\newacronym{UN}{UN}{United Nations}
\newacronym{ALNAP}{ALNAP}{Active Learning Network for Accountability and Performance}
\newacronym{MMS}{MMS}{Moment Magnitude Scale}
\newacronym{MMI}{MMI}{Modified Mercalli Intensity Scale}
\newacronym{ITU}{ITU}{International Telegraph Union}
\newacronym{AP}{AP}{Associated Press}
\newacronym{TCIS}{TCIS}{Tropical Cyclone Intensity Scale}
\newacronym{IMD}{IMD}{India Meteorological Department}
\newacronym{FMS}{FMS}{Fiji Meteorological Service}
\newacronym{UNDRO}{UNDRO}{United Nations Disaster Relief Organization}
\newacronym{JMA}{JMA}{Japan Meteorological Agency}
\newacronym{RSF}{RSF}{Receive, Store and Forward}
\newacronym{OSM}{OSM}{Open Street Map}
\newacronym{PIP}{PIP}{Partners In Health}
\newacronym{STC}{STC}{Store, Carry, Forward}
\newacronym{CRED}{CRED}{Centre for Research on the Epidemiology of Disasters}
\newacronym{RFC}{RFC}{Request For Comment}
\newacronym{UNHCR}{UNHCR}{United Nations High Commissioner for Refugees}
\newacronym{USAF}{USAF}{United States Air Force}
\newacronym{USAFCCT}{USAFCCT}{United States Air Force Combat Control Team}
\newacronym{ATC}{ATC}{Air Traffic Control}
\newacronym{GDACS}{GDACS}{Global Disaster Alert and Coordination System}
\newacronym{INSARAG}{INSARAG}{International Search and Rescue Advisory Group}
\newacronym{UNDAC}{UNDAC}{United Nations Disaster Assessment and Coordination}
\newacronym{NGO}{NGO}{Non-Governmental Organization}
\newacronym{UNO}{UNO}{United Nations Organization}
\newacronym{ECHO}{ECHO}{Directorate-General for European Civil Protection and Humanitarian Aid Operations}
\newacronym{OCHA}{OCHA}{Office for the Coordination of Humanitarian Affairs}
\newacronym{UNISDR}{UNISDR}{United Nations Office for Disaster Risk Reduction}
\newacronym{ETC}{ETC}{Emergency Telecommunications Cluster}
\newacronym{FITTEST}{FITTEST}{Fast IT and Telecommunications Emergency and Support Team}
\newacronym{WFP}{WFP}{World Food Programme}
\newacronym{WHO}{WHO}{World Health Organization}
\newacronym{UNICEF}{UNICEF}{United Nations Children's Fund}
\newacronym{OSOCC}{OSOCC}{On-Site Operations Coordination Centre}
\newacronym{VOSOCC}{VOSOCC}{Virtual On-Site Operations Coordination Centre}
\newacronym{RDC}{RDC}{Reception/Departure Centre}
\newacronym{US}{US}{United States}
\newacronym{SU}{SU}{Soviet Union}
\newacronym{GIS}{GIS}{Geographic Information System}
\newacronym{GUI}{GUI}{Graphical User Interface}
\newacronym{SSHWS}{SSHWS}{Saffir–Simpson Hurricane Wind Scale}
\newacronym{JTWC}{JTWC}{Joint Typhoon Warning Center}
\newacronym{UNSTAMIH}{UNSTAMIH}{United Nations Stabilization Mission In Haiti}
\newacronym{UNHRC}{UNHRC}{United Nations Human Rights Council}
\newacronym{MANET}{MANET}{Mobile Ad-Hoc Network}
\newacronym{NICER}{NICER}{Networked Infrastructureless Cooperation for Emergency Response}
\newacronym{TU}{TU}{Technische Universit\"at}
\newacronym{IDR}{IDR}{International Disaster Response}
\newacronym{DRT}{DRT}{Disaster Response Team}
\newacronym{IDRO}{IDRO}{International Disaster Response Operation}
\newacronym{DRO}{DRO}{Disaster Response Organization}
\newacronym{HDI}{HDI}{Human Development Index}
\newacronym{RWP}{RWP}{Random Waypoint}
\newacronym{POI}{POI}{point of interest}

%% file: introduction.tex

\section{Introduction}
\label{sec:introduction}

Around the globe, we observe a continuous increase in natural disaster occurrences ~\cite{2013LeaningNaturalDisasters}.
When a disaster strikes, the communication infrastructure is often destroyed or unavailable in the immediate aftermath which hinders effective disaster relief work~\cite{Goldstein2010haiti,IRINNews2013haiyan}.

In the humanitarian sector, messaging applications have proved extremely helpful for both organizations as well as individuals as they facilitate coordination and broadcasting information to the public~\cite{ICRC2017}.
Smartphone-based \glspl{DTN} present themselves as an attractive technology for message-based communication in disaster scenarios where the communication infrastructure is disrupted or destroyed as these devices are readily available~\cite{Nishiyama2014}. \glspl{DTN} are facilitated by user mobility as user devices act as data mules to carry messages from $a$ to $b$.
Clearly, the underlying human mobility greatly affects \glspl{DTN} performance as disconnected network partitions are unable to communicate.

Current works on \glspl{DTN} usually conduct simulation-based performance evaluations but are lacking realistic mobility models for large-scale natural disasters.
This lack can be attributed to the unavailability of public mobility traces due to security and privacy concerns of cellular network operators as well as \glspl{DRO}.
In this paper, we approach the problem by creating a disaster scenario mobility model based on expert knowledge.
In particular, we make the following contributions which are relevant for both the opportunistic and ad hoc network as well as the disaster response communities:
\begin{itemize}
	\item We model the mobility of disaster response teams as well as the local population in real large-scale natural disasters. Our model is based on expert knowledge, such as operational reports and conducted interviews, gathered from 126 individuals from 71 \glspl{DRO}.
	\item We characterize our mobility model and compare it with two other widely used models via simulation. In addition, we demonstrate the impact of mobility on \gls{DTN} performance.
	\item We provide an open source implementation~\cite{NatdisCode} of our \emph{Natural Disaster} mobility model as well as two exemplary sce\-narios: the \emph{2010 Haiti earthquake}~\cite{Goldstein2010haiti} and the \emph{2013 Typhoon Haiyan}~\cite{IRINNews2013haiyan}.
\end{itemize}

The rest of the paper is structured as follows:
we first revisit the state-of-the-art of mobility models and \gls{DTN} simulation in \cref{sec:background}.
We explain our approach in \cref{sec:methodology}. Then, we present our mobility model and scenario design in \cref{sec:design} and their implementation in \cref{sec:implementation}.
We analyze and evaluate the impact of our mobility model in \cref{sec:evaluation} and, finally, conclude in \cref{sec:conclusion}.

%% file: background.tex

\section{Related Work}
\label{sec:background}

\begin{table}
\centering
\caption{Related publications on disaster and emergency response mobility models. (\ensuremath{\mathbf{\ast}}) indicates a work that proposes a new model. Our work is printed in bold.}
\label{tab:relatedwork}
\begin{tabulary}{0pt}{@{} crrrl @{}}
	\toprule
	\textsc{Work} & \textsc{Nodes} & \textsc{Area} (m$^2$) & Duration & \textsc{Mobility Model} \\
	\midrule
	\cite{Nelson:2007el} & 100 & 1000 & 25 min & \emph{Event--Role} \ensuremath{\mathbf{\ast}} \\
	\cite{Aschenbruck:2009gq} & 150 & 350 $\times$ 200 & 27.7 h & \emph{Disaster Area} \ensuremath{\mathbf{\ast}} \\	
	\cite{Martin-Campillo2013} & $\leq$ 80 & 700 $\times$ 600 & 10 min & \emph{Disaster Area} \cite{Aschenbruck:2009gq} \\
	\cite{2004AschenbruckHumanMobility} & $\leq$ 200 & $\leq$ 550 $\times$ 500 & 30 min & \emph{Disaster Area} \cite{Aschenbruck:2009gq} \\
	\cite{2012ReinaModellingAssessing} & 200 & 550 $\times$ 500 & 5 min & \emph{Disaster Area} \cite{Aschenbruck:2009gq} \\
	\cite{Krug:2015:ITM:2799371.2799388} & 150 & 350 $\times$ 200 & 1.1 h & \emph{Disaster Area} \cite{Aschenbruck:2009gq} \\
	\cite{6918914} & 100 & 6000 $\times$ 5000 & 35 h & \emph{Search \& Rescue} \ensuremath{\mathbf{\ast}} \\
	\cite{5429249} & 234 & N/A & 4 days & \emph{Natural Disaster} \ensuremath{\mathbf{\ast}} \\
	\textbf{---} & \textbf{500} & \textbf{5000 $\times$ 7000} & \textbf{7 days} & \textbf{\emph{Natural Disaster} \ensuremath{\mathbf{\ast}}} \\
	\bottomrule
\end{tabulary}
\end{table}


Generic mobility models such as \gls{RWP} have been used to study the performance of mobile wireless networks such as \glspl{DTN}. However, these models do not capture the non-randomness of human mobility and, therefore, produce questionable results when trying to understand network performance in realistic scenarios.
To alleviate this problem, multiple works (see \cref{tab:relatedwork}) have sought to create domain-specific models:
\cite{Aschenbruck:2009gq} proposes \emph{Disaster Area} which models disaster scenarios based on traces collected at a fire fighter manoeuvre.
This model has been used multiple times to evaluate performance of mobile networks, e.\,g., \cite{Martin-Campillo2013, 2012ReinaModellingAssessing}.
\cite{6918914} proposes a micro-movement pattern generation framework for search-and-rescue missions.
\cite{Krug:2015:ITM:2799371.2799388} states the need for expert knowledge to create realistic scenarios.
\cite{5429249} is closest related to our work as they attempt to model a large-scale post-disaster scenario. However, they claim that ``no documents/reports happen to describe the disaster operation in as detail as required to reproduce the scene by a simulator.''
In our work, we tackle precisely this problem and reverse-engineer human mobility at a granularity-level suitable for a simulator by using information from public and non-public documents and by conducting interviews with various \gls{IDR} experts.


%% file: methodology.tex

\section{Methodology}
\label{sec:methodology}

\begin{figure}[b]
	\centering
	\includegraphics{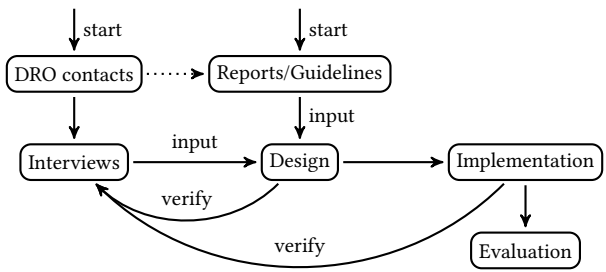}
	\caption{Our process for information gathering, model design and implementation, verification, and evaluation.}
	\label{fig:methodology}
\end{figure}

GPS mobility traces are the ``holy grail'' for creating accurate mobility models.
Unfortunately, such ground truth data during and post disaster especially from \glspl{DRT} are not (publicly) available.
This is due to the plain lack of records, strict data privacy laws, mobile network operators protecting their valuable business assets, as well as security concerns of \glspl{DRO} who do not wish to disclose sensitive location information of their staff, especially when operating in armed-conflict areas.

As a consequence, we follow an alternative approach:
we base our mobility model on ``soft'' data, in particular, expert knowledge of various \glspl{DRO}.
In order for this method to yield acceptable results, we need to ensure that
\begin{inparaenum}
	\item the model input is sufficiently fine-grained for extracting mobility patterns,
	\item the model is verified by a group of experts.
\end{inparaenum}

Unfortunately, documentation of the events during a disaster is often incomplete:
media coverage of high-profile disaster is vast but articles typically lack technical details.
Also, first responders and disaster relief workers are supposed to create daily reports but can often invest only minimal time as they prioritize actual life-saving work. This often leads incomplete or imprecise documentation unusable for our purposes.
To understand \gls{IDR} and eventually derive our mobility model, we gathered information from
\begin{enumerate}
	\item \gls{DRO} policies, guidelines, strategies and fieldbooks partly gathered from online disaster response platforms, and
	\item individual (on-site) interviews with disaster relief experts.
\end{enumerate}
Our approach is visualized in \cref{fig:methodology}.
%
During the process, we contacted 298 and received feedback from 126 \gls{IDR} experts (42\,\% response rate).
From the responders, we received additional information about their humanitarian activities, past mission reports, and pointers to other experts and contacts better suited for replying to our inquiry.
In addition, we received access to exclusive online platforms and forums such as the \gls{VOSOCC} of the UN \gls{OCHA}, reliefweb.int, and humanitarianresponse.info. They contain field handbooks and guidelines for in-field operations which are not publicly accessible.
Furthermore, we were able to conduct 15 on-site interviews with \gls{IDR} experts and former disaster relief workers.
Those experts were also involved in the verification loop from design to implementation.
%

%% file: design.tex

\begin{table*}
	\newcommand{\myast}{\ensuremath{\mathbf{\ast}}}
	\newcommand{\low}{\myast}
	\newcommand{\medium}{\myast\myast}
	\newcommand{\high}{\myast\myast\myast}
	\centering
	\caption{Large-scale natural disasters in the last ten years. Ratings range from low (\low) to high (\high). Our scenarios are printed in \textbf{bold}. We selected Typhoon Haiyan as the running example for this paper.}
	\label{tab:inspectedDisasters}
	\begin{tabulary}{0pt}{@{}lclrrrrcc@{}}
		\toprule
		\textsc{Disaster} & \textsc{Year} & \textsc{Country} & \textsc{Killed} & \textsc{Injured} & \textsc{Displaced} & \textsc{Area (\textnormal{km$^2$})} & \textsc{Response} & \textsc{Information} \\
		\midrule
		Nepal earthquake & 2015 & Nepal & 9\,000 & 22\,000 & 2\,600\,000 & 3\,610 & \high & \high \\
		Cyclone Pam & 2015 & Vanuatu & 24 & 30 & 3\,300 & 12\,190 & \medium & \medium \\
		Ludian earthquake & 2014 & China & 617 & 2\,400 & 229\,700 & 1\,487 & \low & \low \\
		\textbf{Typhoon Haiyan} & \textbf{2013} & \textbf{Philippines} & \textbf{6\,300} & \textbf{28\,689} & \textbf{6\,000\,000} & \textbf{71\,503} & \high & \high \\
		Christchurch earthquake & 2011 & New Zealand & 185 & 2\,000 & --- & 1\,426 & \medium & \high \\
		East Africa drought & 2011 & East Africa & 260\,000 & --- & 5\,720\,000 & 2\,346\,466 & \high & \high \\
		Tropical storm Washi & 2011 & Philippines & 1\,292 & 2\,002 & 430\,900 & 104\,530 & \medium & \low \\
		Tohoku earthquake & 2011 & Japan & 15\,894 & 6\,152 & 340\,000 & 83\,955 & \medium & \high \\
		\textbf{Haiti earthquake} & \textbf{2010} & \textbf{Haiti} & \textbf{316\,000} & \textbf{300\,000} & \textbf{895\,000} & \textbf{27\,750} & \high & \high \\
		Afghanistan blizzard & 2008 & Afghanistan & 926 & 100 & --- & 652\,864 & \low & \low \\
		Sichuan earthquake & 2008 & China & 69\,195 & 374\,643 & 5\,000\,000 & 485\,000 & \medium & \high \\
		Cyclone Nargis & 2008 & Myanmar & 138\,000 & --- & 800\,000 & 676\,578 & \medium & \medium \\
		\bottomrule& & 
	\end{tabulary}
\end{table*}

\section{Design}
\label{sec:design}
In this section, we review past candidate events which qualify for our methodology. We then provide a detailed time-line of events for one recent high-profile disaster: 2013 Typhoon Haiyan. The same information for the 2010 Haiti earthquake is available in \cite{Schons2017} but was omitted here for brevity.
From the description and a number of policy documents and guidelines provided by \glspl{DRO}, we extract structured elements such as the different actors and their activities which we finally pour into a novel generic mobility model.

\subsection{Scenario Selection Criteria}
\label{sec:scenario-selection}

\Cref{tab:inspectedDisasters} lists recent natural disasters and their scale.
We found that the following six factors are key for the creation of scenarios and models for \gls{DTN} research in the area of natural disasters:
\begin{inparaenum}
	\item the number of \emph{affected people} and 
	\item the size of the affected \emph{area} both define the scope of the disaster;
	\item \emph{geographic conditions} are important as impassable natural obstacles such as rivers or lakes would partition the network and render a \gls{DTN} deployment unusable;
	\item the availability of \emph{infrastructure} (bridges, airports, power plants, cellular network, etc.) after the disaster; and
	\item strong international \emph{response} by \glspl{DRO} as they are the sources for 
	\item sufficient, fine-grained \emph{information} on the event.
\end{inparaenum}
The latter is the most important factor for us as we base our model and scenario design solely on written and oral reports.
Note that we did not include all presented factors in \cref{tab:inspectedDisasters} due to space reasons, but provide an extended version in \cite{Schons2017}.

\subsection{2013 Typhoon Haiyan}
\label{sec:scenario-haiyan}

We chose to re-create human mobility of the 2013 Typhoon Haiyan as media coverage and response was high, and communication infrastructure was dysfunctional during the first days. In this section, we will give an overview of the disaster and describe the situation in detail during the first week (days zero to six).
The following information was gathered through interviews\footnote{Interview notes are available upon request.} with multiple first responder eyewitnesses from different relief organizations, as well as press articles about the relief efforts.

\subsubsection{Summary}
\begin{figure}
	\centering
	\includegraphics[width=0.8\linewidth]{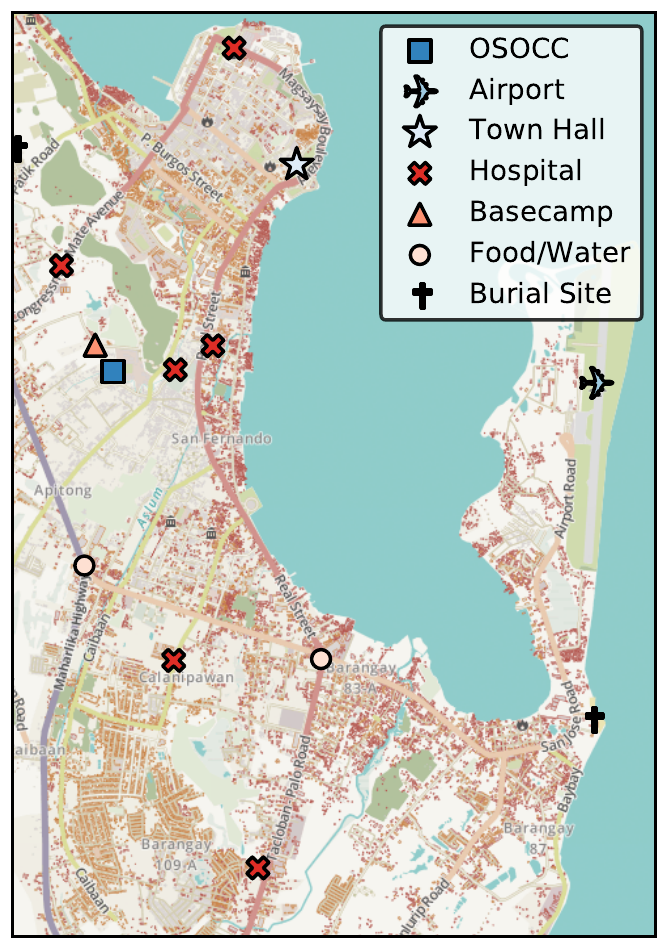}
	\caption{City of Tacloban with locations of \gls{IDR}-relevant sites in the early aftermath of 2013 Typhoon Haiyan.}
	\label{fig:tacloban-map}
\end{figure}

Typhoon Haiyan lasted from November 3rd to 11th, 2013 and was one of the strongest tropical cyclones ever recorded \cite{NPRHaiyanDamage}. Even though typhoon Haiyan had devastating effects on large portions of Southeast Asia, for the purpose of this work we will focus on the aftermath of November 8th when Typhoon Haiyan hit the Philippines at 04:40 local time. Haiyan was the deadliest and most damaging Philippine typhoon on record and left more than one million houses partially or totally damaged, killing at least 6\,300 people and leaving numerous injured and homeless \cite{NDRRMCSitUpdt}. Typhoon Haiyan was ranked as a category 5 typhoon, the highest category by the definition of the \gls{SSHWS}, implying that \enquote{catastrophic damage will occur} and \enquote{most of the area will be uninhabitable for weeks or months.}
After the storm had passed, widespread damage became visible with power lines cut off, roads blocked by fallen debris, and trees and buildings collapsed under the strong winds \cite{DisasterCharterActivation}. A 2013 preliminary estimate \cite{NDRRMCFinalReportYolanda} calculated the total damage related to typhoon Haiyan to be around 2.86 billion US\$.
Figure \ref{fig:tacloban-map} depicts the city of Tacloban and its surroundings, which were severely hit by Typhoon Haiyan, and is intended to provide visual guidance to the reader for a better understanding of the upcoming sections. 

%
Right after the disaster had hit the Philippines, many officials concluded that even though early warnings had been issued to the population of potentially affected areas, only few people actually evacuated. This was likely related to the high number of smaller typhoons the Philippines experience every year, which led to the population underestimating the severity of the coming typhoon. Warnings were broadcast on TV and radio, but went largely unheeded.
%
The typhoon was accompanied by the biggest storm surge ever experienced within the Philippines, reportedly reaching between four to six meters in height \cite{TelegraphHaiyanPhilippines}. This resulted in fast-rising tides and surge water brought by the typhoon, which led to many additional fatalities \cite{DailymotionMegastormDocumentation}.

\subsubsection{Day 0}
In the aftermath of the typhoon, the immediate arrival of \glspl{DRT} was hindered by the severe damage the airport had sustained. Therefore, many \glspl{DRT} were rerouted to the airport of the nearby island of Cebu, which was still operational. From there, they had to travel to Tacloban using other means, which took around 8 hours. This slowed down the initial arrival of \glspl{DRT} considerably. Despite the damage, a number of airplanes and helicopters managed to land in Tacloban, delivering aid and personnel. The newly arrived \glspl{DRT} registered at the \gls{RDC} before proceeding into the city to reach the \gls{OSOCC}.
There were also a number of storm chasers and typhoon experts already on the ground when Haiyan made landfall. In search of scientific evidence, they moved through the city before, during and after the typhoon. (For our mobility model, we will assume that they are distributed randomly throughout the city.)
The local population spent the storm in their homes or shelters. After the storm had passed, eyewitnesses report that most of the population seemed to wander around the perimeter of their home or shelter area, overwhelmed by the destruction caused by the storm, or trying to help their neighbors. (For our mobility model, we will assume their locations to be randomly spread throughout the city.)
%
At the end of the day, \glspl{DRT} and experts alike returned to their respective sleeping spots.

\subsubsection{Day 1 to 3}
On the first day after the disaster, the US Navy deployed radar equipment to the Tacloban airport, allowing it to be re-opened in short order. This allowed a greater number of \glspl{DRT} and relief goods to arrive close to the city, leading to more travel between the airport and the city. Arriving \glspl{DRT} reported at the \gls{RDC} and then proceeded to the \gls{OSOCC}, which was located in the city center. Afterwards, they traveled to the base camp of their organization and began their relief efforts.
The initial focus of most \glspl{DRT} lies with \glspl{SRO}, as the likelihood of finding survivors drops with every hour. \glspl{USRT} have a large number of different search strategies. (For the purpose of the movement model, we implement a basic strategy: a start area is chosen, and the team searches every house along that street before spreading out to the next street in the direct neighborhood.)
Eyewitnesses indicate that during the first three days, the civilian population was mostly concerned with search for food and the rebuilding of their homes or temporary shelters. Healthy survivors assisted with local search and rescue efforts and cleared the street from debris, staying in the general area of their homes or shelters. Injured survivors were transported to one of the partially operational city hospitals, which were quickly filled to capacity.
As the government still remained in office, the relief organizations had to coordinate with country officials. This led to some traffic between the city hall and \gls{OSOCC}. The \gls{UN} rented a hotel as a temporary office location for the duration of the rebuilding efforts.\footnote{We were asked not to publish the location of the hotel, so we randomly chose a hotel in the city for the purposes of the simulation.} All \glspl{DRT} continued to return to their respective base camps for the night.


\subsubsection{Day 4 to 6}
\glspl{SRO} were reduced and came to a stop at the end of the first week, as the chance of finding survivors drops significantly after several days. After about one week, most of the \glspl{USRT} are replaced by teams specialized in other forms of disaster relief. The distribution of food and clean water remained a challenge, as the infrastructure was severely damaged by the typhoon. Much of the local population had to collect food and water from distribution points around the city on a daily basis.
The local population assisted \glspl{DRT} in clearing the roads of debris to allow trucks to pass through them. The government also ordered the removal of dead bodies to prevent the spread of diseases. The US Marines started flying out injured locals and exhausted relief workers alike. At the same time, new \glspl{DRT} were still arriving via the airport, following the same procedure as previously described.

\subsubsection{After Week 1}
\label{subsec:endofhaiyan}
We stop our description of the events after the first week. At this time, locals had slowly begun to resume a semblance of their normal lives, and the scenario gradually transforms into a scenario of daily routine, where a dedicated mobility model no longer applies.
%
For the same reason, we did not consider that most \glspl{DRO} have a rotation mechanism for their employees, which prevents them from becoming mentally and physically exhausted from the demanding work in disaster relief.

\begin{figure*}
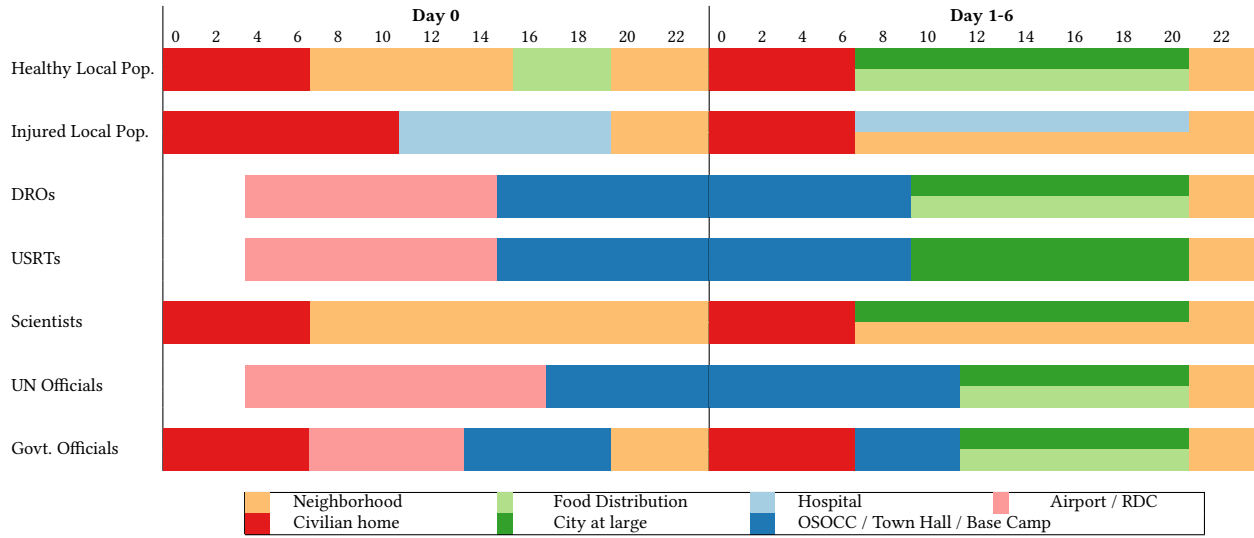

	\centering
	\footnotesize
	\setlength\tabcolsep{3pt} 
\begin{tabulary}{\textwidth}{lllllllllllllllllllllllllllllllllllllllllllllllll}
	\multicolumn{1}{l|}{} & \multicolumn{24}{c|}{\textbf{Day 0}} & \multicolumn{24}{c}{\textbf{Day 1-6}} \\
	\multicolumn{1}{l|}{} & 0 &  & 2 &  & 4 &  & 6 &  & 8 &  & 10 &  & 12 &  & 14 &  & 16 &  & 18 &  & 20 &  & 22 & \multicolumn{1}{l|}{} & 0 & & 2 &  & 4 &  & 6 &  & 8 &  & 10 &  & 12 &  & 14 &  & 16 &  & 18 &  & 20 &  & 22 &  \\
	\multicolumn{1}{l|}{} & \multicolumn{7}{l}{\cellcolor[HTML]{E31A1C}} & \multicolumn{4}{l}{\cellcolor[HTML]{FDBF6F}} & \multicolumn{5}{l}{\cellcolor[HTML]{FDBF6F}} & \multicolumn{4}{l}{\cellcolor[HTML]{B2DF8A}} & \multicolumn{4}{l|}{\cellcolor[HTML]{FDBF6F}} & \multicolumn{7}{l}{\cellcolor[HTML]{E31A1C}} & \multicolumn{14}{l}{\cellcolor[HTML]{33A02C}} & \multicolumn{3}{l}{\cellcolor[HTML]{FDBF6F}} \\
	\multicolumn{1}{l|}{\multirow{-2}{*}{Healthy Local Pop.}} & \multicolumn{7}{l}{\multirow{-2}{*}{\cellcolor[HTML]{E31A1C}}} & \multicolumn{4}{l}{\multirow{-2}{*}{\cellcolor[HTML]{FDBF6F}}} & \multicolumn{5}{l}{\cellcolor[HTML]{FDBF6F}} & \multicolumn{4}{l}{\multirow{-2}{*}{\cellcolor[HTML]{B2DF8A}}} & \multicolumn{4}{l|}{\multirow{-2}{*}{\cellcolor[HTML]{FDBF6F}}} & \multicolumn{7}{l}{\multirow{-2}{*}{\cellcolor[HTML]{E31A1C}}} & \multicolumn{14}{l}{\cellcolor[HTML]{B2DF8A}} & \multicolumn{3}{l}{\multirow{-2}{*}{\cellcolor[HTML]{FDBF6F}}} \\
	\multicolumn{1}{l|}{} &  &  &  &  &  &  &  &  &  &  &  &  &  &  &  &  &  &  &  &  &  &  &  & \multicolumn{1}{l|}{} &  &  &  &  &  &  &  &  &  &  &  &  &  &  &  &  &  &  &  &  &  &  &  &  \\
	\multicolumn{1}{l|}{} & \multicolumn{7}{l}{\cellcolor[HTML]{E31A1C}} & \multicolumn{4}{l}{\cellcolor[HTML]{E31A1C}} & \multicolumn{9}{l}{\cellcolor[HTML]{A6CEE3}} & \multicolumn{4}{l|}{\cellcolor[HTML]{FDBF6F}} & \multicolumn{7}{l}{\cellcolor[HTML]{E31A1C}} & \multicolumn{14}{l}{\cellcolor[HTML]{A6CEE3}} & \multicolumn{3}{l}{\cellcolor[HTML]{FDBF6F}} \\
	\multicolumn{1}{l|}{\multirow{-2}{*}{Injured Local Pop.}} & \multicolumn{7}{l}{\multirow{-2}{*}{\cellcolor[HTML]{E31A1C}}} & \multicolumn{4}{l}{\multirow{-2}{*}{\cellcolor[HTML]{E31A1C}}} & \multicolumn{9}{l}{\cellcolor[HTML]{A6CEE3}} & \multicolumn{4}{l|}{\multirow{-2}{*}{\cellcolor[HTML]{FDBF6F}}} & \multicolumn{7}{l}{\multirow{-2}{*}{\cellcolor[HTML]{E31A1C}}} & \multicolumn{14}{l}{\cellcolor[HTML]{FDBF6F}} & \multicolumn{3}{l}{\multirow{-2}{*}{\cellcolor[HTML]{FDBF6F}}} \\
	\multicolumn{1}{l|}{} &  &  &  &  &  &  &  &  &  &  &  &  &  &  &  &  &  &  &  &  &  &  &  & \multicolumn{1}{l|}{} &  &  &  &  &  &  &  &  &  &  &  &  &  &  &  &  &  &  &  &  &  &  &  &  \\
	\multicolumn{1}{l|}{} &  &  &  &  & \multicolumn{11}{l}{\cellcolor[HTML]{FB9A99}} & \multicolumn{2}{l}{\cellcolor[HTML]{1F78B4}} & \multicolumn{3}{l}{\cellcolor[HTML]{1F78B4}} & \multicolumn{4}{l|}{\cellcolor[HTML]{1F78B4}} & \multicolumn{7}{l}{\cellcolor[HTML]{1F78B4}} & \multicolumn{3}{l}{\cellcolor[HTML]{1F78B4}} & \multicolumn{11}{l}{\cellcolor[HTML]{33A02C}} & \multicolumn{3}{l}{\cellcolor[HTML]{FDBF6F}} \\
	\multicolumn{1}{l|}{\multirow{-2}{*}{DROs}} &  &  &  &  & \multicolumn{11}{l}{\multirow{-2}{*}{\cellcolor[HTML]{FB9A99}}} & \multicolumn{2}{l}{\multirow{-2}{*}{\cellcolor[HTML]{1F78B4}}} & \multicolumn{3}{l}{\multirow{-2}{*}{\cellcolor[HTML]{1F78B4}}} & \multicolumn{4}{l|}{\multirow{-2}{*}{\cellcolor[HTML]{1F78B4}}} & \multicolumn{7}{l}{\multirow{-2}{*}{\cellcolor[HTML]{1F78B4}}} & \multicolumn{3}{l}{\multirow{-2}{*}{\cellcolor[HTML]{1F78B4}}} & \multicolumn{11}{l}{\cellcolor[HTML]{B2DF8A}} & \multicolumn{3}{l}{\multirow{-2}{*}{\cellcolor[HTML]{FDBF6F}}} \\
	\multicolumn{1}{l|}{} &  &  &  &  &  &  &  &  &  &  &  &  &  &  &  &  &  &  &  &  &  &  &  & \multicolumn{1}{l|}{} &  &  &  &  &  &  &  &  &  &  &  &  &  &  &  &  &  &  &  &  &  &  &  &  \\
	\multicolumn{1}{l|}{} &  &  &  &  & \multicolumn{11}{l}{\cellcolor[HTML]{FB9A99}} & \multicolumn{2}{l}{\cellcolor[HTML]{1F78B4}} & \multicolumn{3}{l}{\cellcolor[HTML]{1F78B4}} & \multicolumn{4}{l|}{\cellcolor[HTML]{1F78B4}} & \multicolumn{7}{l}{\cellcolor[HTML]{1F78B4}} & \multicolumn{3}{l}{\cellcolor[HTML]{1F78B4}} & \multicolumn{11}{l}{\cellcolor[HTML]{33A02C}} & \multicolumn{3}{l}{\cellcolor[HTML]{FDBF6F}} \\
	\multicolumn{1}{l|}{\multirow{-2}{*}{USRTs}} &  &  &  &  & \multicolumn{11}{l}{\multirow{-2}{*}{\cellcolor[HTML]{FB9A99}}} & \multicolumn{2}{l}{\multirow{-2}{*}{\cellcolor[HTML]{1F78B4}}} & \multicolumn{3}{l}{\multirow{-2}{*}{\cellcolor[HTML]{1F78B4}}} & \multicolumn{4}{l|}{\multirow{-2}{*}{\cellcolor[HTML]{1F78B4}}} & \multicolumn{7}{l}{\multirow{-2}{*}{\cellcolor[HTML]{1F78B4}}} & \multicolumn{3}{l}{\multirow{-2}{*}{\cellcolor[HTML]{1F78B4}}} & \multicolumn{11}{l}{\multirow{-2}{*}{\cellcolor[HTML]{33A02C}}} & \multicolumn{3}{l}{\multirow{-2}{*}{\cellcolor[HTML]{FDBF6F}}} \\
	\multicolumn{1}{l|}{} &  &  &  &  &  &  &  &  &  &  &  &  &  &  &  &  &  &  &  &  &  &  &  & \multicolumn{1}{l|}{} &  &  &  &  &  &  &  &  &  &  &  &  &  &  &  &  &  &  &  &  &  &  &  &  \\
	\multicolumn{1}{l|}{} & \multicolumn{7}{l}{\cellcolor[HTML]{E31A1C}} & \multicolumn{13}{l}{\cellcolor[HTML]{FDBF6F}} & \multicolumn{4}{l|}{\cellcolor[HTML]{FDBF6F}} & \multicolumn{7}{l}{\cellcolor[HTML]{E31A1C}} & \multicolumn{14}{l}{\cellcolor[HTML]{33A02C}} & \multicolumn{3}{l}{\cellcolor[HTML]{FDBF6F}} \\
	\multicolumn{1}{l|}{\multirow{-2}{*}{Scientists}} & \multicolumn{7}{l}{\multirow{-2}{*}{\cellcolor[HTML]{E31A1C}}} & \multicolumn{13}{l}{\multirow{-2}{*}{\cellcolor[HTML]{FDBF6F}}} & \multicolumn{4}{l|}{\multirow{-2}{*}{\cellcolor[HTML]{FDBF6F}}} & \multicolumn{7}{l}{\multirow{-2}{*}{\cellcolor[HTML]{E31A1C}}} & \multicolumn{14}{l}{\cellcolor[HTML]{FDBF6F}} & \multicolumn{3}{l}{\multirow{-2}{*}{\cellcolor[HTML]{FDBF6F}}} \\
	\multicolumn{1}{l|}{} &  &  &  &  &  &  &  &  &  &  &  &  &  &  &  &  &  &  &  &  &  &  &  & \multicolumn{1}{l|}{} &  &  &  &  &  &  &  &  &  &  &  &  &  &  &  &  &  &  &  &  &  &  &  &  \\
	\multicolumn{1}{l|}{} &  &  &  &  & \multicolumn{11}{l}{\cellcolor[HTML]{FB9A99}} & \multicolumn{2}{l}{\cellcolor[HTML]{FB9A99}} & \multicolumn{3}{l}{\cellcolor[HTML]{1F78B4}} & \multicolumn{4}{l|}{\cellcolor[HTML]{1F78B4}} & \multicolumn{7}{l}{\cellcolor[HTML]{1F78B4}} & \multicolumn{5}{l}{\cellcolor[HTML]{1F78B4}} & \multicolumn{9}{l}{\cellcolor[HTML]{33A02C}} & \multicolumn{3}{l}{\cellcolor[HTML]{FDBF6F}} \\
	\multicolumn{1}{l|}{\multirow{-2}{*}{UN Officials}} &  &  &  &  & \multicolumn{11}{l}{\multirow{-2}{*}{\cellcolor[HTML]{FB9A99}}} & \multicolumn{2}{l}{\multirow{-2}{*}{\cellcolor[HTML]{FB9A99}}} & \multicolumn{3}{l}{\multirow{-2}{*}{\cellcolor[HTML]{1F78B4}}} & \multicolumn{4}{l|}{\multirow{-2}{*}{\cellcolor[HTML]{1F78B4}}} & \multicolumn{7}{l}{\multirow{-2}{*}{\cellcolor[HTML]{1F78B4}}} & \multicolumn{5}{l}{\multirow{-2}{*}{\cellcolor[HTML]{1F78B4}}} & \multicolumn{9}{l}{\cellcolor[HTML]{B2DF8A}} & \multicolumn{3}{l}{\multirow{-2}{*}{\cellcolor[HTML]{FDBF6F}}} \\
	\multicolumn{1}{l|}{} &  &  &  &  &  &  &  &  &  &  &  &  &  &  &  &  &  &  &  &  &  &  &  & \multicolumn{1}{l|}{} &  &  &  &  &  &  &  &  &  &  &  &  &  &  &  &  &  &  &  &  &  &  &  &  \\
	\multicolumn{1}{l|}{} & \multicolumn{7}{l}{\cellcolor[HTML]{E31A1C}} & \multicolumn{7}{l}{\cellcolor[HTML]{FB9A99}} & \multicolumn{4}{l}{\cellcolor[HTML]{1F78B4}} & \multicolumn{2}{l}{\cellcolor[HTML]{1F78B4}} & \multicolumn{4}{l|}{\cellcolor[HTML]{FDBF6F}} & \multicolumn{7}{l}{\cellcolor[HTML]{E31A1C}} & \multicolumn{5}{l}{\cellcolor[HTML]{1F78B4}} & \multicolumn{9}{l}{\cellcolor[HTML]{33A02C}} & \multicolumn{3}{l}{\cellcolor[HTML]{FDBF6F}} \\
	\multicolumn{1}{l|}{\multirow{-2}{*}{Govt. Officials}} & \multicolumn{7}{l}{\multirow{-2}{*}{\cellcolor[HTML]{E31A1C}}} & \multicolumn{7}{l}{\multirow{-2}{*}{\cellcolor[HTML]{FB9A99}}} & \multicolumn{4}{l}{\multirow{-2}{*}{\cellcolor[HTML]{1F78B4}}} & \multicolumn{2}{l}{\multirow{-2}{*}{\cellcolor[HTML]{1F78B4}}} & \multicolumn{4}{l|}{\multirow{-2}{*}{\cellcolor[HTML]{FDBF6F}}} & \multicolumn{7}{l}{\multirow{-2}{*}{\cellcolor[HTML]{E31A1C}}} & \multicolumn{5}{l}{\multirow{-2}{*}{\cellcolor[HTML]{1F78B4}}} & \multicolumn{9}{l}{\cellcolor[HTML]{B2DF8A}} & \multicolumn{3}{l}{\multirow{-2}{*}{\cellcolor[HTML]{FDBF6F}}} \\
 &  &  &  &  &  &  &  &  &  &  &  &  &  &  &  &  &  &  &  &  &  &  &  &  &  &  &  &  &  &  &  &  &  &  &  &  &  &  &  &  &  &  &  &  &  &  &  &  \\ \cline{6-47}
 &  &  &  & \multicolumn{1}{l|}{} &   \cellcolor[HTML]{FDBF6F} &  & \multicolumn{7}{l}{Neighborhood} &  &  & \cellcolor[HTML]{B2DF8A} &  & \multicolumn{7}{l}{Food Distribution}  &  &  & \cellcolor[HTML]{A6CEE3} &  & \multicolumn{7}{l}{Hospital} &  &  & \cellcolor[HTML]{FB9A99} &  & \multicolumn{7}{l|}{Airport / RDC} &  &  \\
 &  &  &  & \multicolumn{1}{l|}{} & \cellcolor[HTML]{E31A1C} &  & \multicolumn{7}{l}{Civilian home} &  &  & \cellcolor[HTML]{33A02C} &  & \multicolumn{7}{l}{City at large} &  &  & \cellcolor[HTML]{1F78B4} &  & \multicolumn{18}{l|}{OSOCC / Town Hall / Base Camp} &  &  \\ \cline{6-47}
 &  &  &  &  &  &  &  &  &  &  &  &  &  &  &  &  &  &  &  &  &  &  &  &  &  &  &  &  &  &  &  &  &  &  &  &  &  &  &  &  &  &  &  &  &  &  &  & 
	\end{tabulary}
	\vspace{-0.3cm}
	\caption{Activity-based location of different groups over time. Time values are indicative, i.\,e., activities do not start exactly at the same time but with a random jitter of one or two hours.}
	\label{fig:pois}
\end{figure*}

\subsection{Natural Disaster Mobility Model}
\label{sec:movementmodeldescriptionanddesign}

Based on the information gathered for specific disasters, we extract recurring behavioral patterns of the various entities involved in disaster relief work.
To this end, we define \emph{roles} and role-specific \emph{activities}, and also consider movement \emph{speed}.

\subsubsection{Roles}
We identified the main stakeholders in natural disaster-struck areas (scenario-independent) and defined the following seven roles with distinct behavioral patterns:
\begin{inparaenum}
	\item healthy local population,
	\item injured local population,
	\item \glsreset{DRT} \glspl{DRT} from \glspl{DRO},
	\item dedicated \glsreset{USRT} \glspl{USRT},
	\item scientists (storm chasers, typhoon experts, etc.),
	\item \gls{UN} officials, and
	\item government officials.
\end{inparaenum} 

\subsubsection{Activities}
To create a model, we further need to define activities that regularly occur in disaster areas that can be attributed to specific roles. 
\Cref{fig:pois} summarizes the various activities by showing the coarse locations of each role during the course of the first week.
What follows is a detailed but non-exhaustive description of important activities that have been identified during the interviews with \gls{IDR} experts.

\emph{Activities applying to everyone.}
In the evening, everyone goes to their respective base camp, home, or shelter to sleep.
Those arriving via the airport (e.\,g., \glspl{DRO}), at the day of arrival, first go to the \gls{RDC} for registration, then visit the \gls{OSOCC}, and finally set up the base camp or sleeping place.

\emph{Activities applying to \glspl{DRO} and \glspl{DRT}.}
After arriving and registering at the airport, they go to the \gls{OSOCC} or the town hall for a situation briefing and then start to help the affected population with one of the following activities:
\begin{inparaenum}[($i$)]
	\item collect dead bodies and organize burials;
	\item walk the main streets of the city and clean streets from debris, such that supplies can be delivered; or
	\item go to food and water distribution centers to serve the locals until the end of the day.
\end{inparaenum}
Besides that, they regularly visit the \gls{UN} hotel, the \gls{OSOCC} site, the base camp, or town hall to meet with officials and other \glspl{DRO}.

\emph{Activities applying to \gls{UN} and government officials.}
Officials regularly (at least daily) visit the \gls{OSOCC}, the town hall or base camp, for a situation briefing and meet other officials and \glspl{DRO}.
During the day, they perform reconnaissance missions to get an situation overview such as infrastructural damage. This information is used to provision help.
Also, they organize the disaster relief efforts with other officials and \glspl{DRO} such as setting up food and water distribution spots, organizing burials, etc.

\emph{Activities applying to scientists.}
Within the first two to three days, they collect scientific evidence from the disaster site before the cleaning of the rubble and debris starts.
When their job is done, they either leave the area via the airport, or
decide to volunteer and help the \glspl{DRO}.

\emph{Activities applying to \gls{USRT}.}
After arriving and registering at the airport, they go to the \gls{OSOCC} for a situation briefing and \gls{SRO} planning.
When starting an \gls{SRO}, \glspl{USRT} go to the chosen location in the morning and then search every house in that street. When done, they repeat with next street in the direct neighborhood.
Usually, \glspl{SRO} are stopped after one week as the chances to find survivors diminish, and \glspl{USRT} fly back home.

\emph{Activities applying to healthy local population.}
According to eye witnesses, most locals stay at home or try to find friends and family members within the immediate surroundings after the disaster has struck.
Later on the first day, they stay in the proximity of their homes to assess the damage and to help their neighbors.
Then, they start to look for food and water, for example, at distribution centers where they will return on a daily basis.
The rest of the day, they either volunteer %
\begin{inparaenum}[]
	\item for cleaning operations (we model this by slowly roaming around the city), or
	\item as replacement of security personnel to patrol the area.
\end{inparaenum}

\emph{Activities applying to injured local population.}
Depending on the severeness, the injured stay at home if they are unable to move, or try to go to the closest hospital as soon as possible.
Upon arrival, they stay at the hospital if its capacity is not exhausted or leave to find another one otherwise.

\subsubsection{Typhoon Haiyan Arrival Times}
Most scientists, particularly storm chasers and typhoon experts were already present prior to the occurrence of the disaster. Furthermore, government officials were already present prior to the disaster, while \gls{UN} officials arrived after the disaster had struck. \glspl{DRO} and \glspl{USRT} mainly arrived via the airport during the first days after the disaster as they were delayed due to the damaged airport. However, some had been prepositioned to locations close to the affected area and arrived on day zero.

\subsubsection{Movement Speed}
For the purpose of this work, we consider walking the only viable form of movement. This is due to the fact that within disaster-struck areas, debris, flood/surge water, or earthquakes often render streets completely unusable for cars and trucks until they can be repaired.
This means that all entities move approximately at walking speed. Nevertheless, there are slight differences depending on the entity's role. For example, injured individuals and heavily equipped \glspl{USRT} will be slower than normal.
Note that our model could be extended to support faster-traveling vehicles, but would require additional scenario-specific information currently unavailable to us, such as who has access to vehicles and which streets are passable.

%% file: implementation.tex

\section{Implementation}
\label{sec:implementation}

In this section, we briefly describe the implementation of our generic Natural Disaster \emph{mobility model} and the workflow for generating specific \emph{scenarios}.
This is a trade-off between code re-usability (reoccurring behavioral patterns of certain groups such as \glspl{DRO} and the local population) and acknowledging the uniqueness of every disaster (terrain, streets, \glspl{POI}, arrival times, population density, etc.)

\begin{figure}
	\centering
	\includegraphics{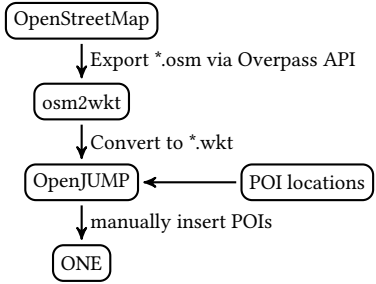}
	\caption{Technical workflow for generating a \emph{scenario}'s map data for our Natural Disaster mobility model.}
	\label{fig:map-one}
\end{figure}

\subsubsection{Mobility Model}
We implement our mobility model directly as a module for the \gls{ONE} \cite{keranen2009theone} as we are concerned with \gls{DTN} simulation.
Our model implements the different roles and activities as defined in \cref{sec:design}. These roles and activities have been observed in different disasters (\cref{tab:inspectedDisasters}) and as such are generally independent of the scenario instance.
Nevertheless, our model can only operate on a specific \emph{scenario}.

\subsubsection{Scenario}
A \emph{scenario} describes the prevailing conditions of a specific post-disaster situation. In particular, the scenario comprises the number of actors as well as a street map including different \glspl{POI} of the affected area.
We generate the map data using a number of tools shown in \cref{fig:map-one}: we start with exporting OpenStreetMap data and converting it to the Well-Known Text (.wkt) file format using \emph{osm2wkt} \cite{osm2wkt}. Finally, we use \emph{OpenJUMP} to add \gls{POI} locations, such as the OSOCC, airports, and hospitals as shown in \cref{fig:tacloban-map}. The .wkt files can then be processed by our mobility model implemented in the \gls{ONE}.
In this paper, we only focus on the area around the city of Tacloban which was severely hit by Typhoon Haiyan. However, we have also implemented a scenario for the 2010 Haiti earthquake (Port-au-Prince and surroundings) which is discussed in \cite{Schons2017} and included in our source code \cite{NatdisCode}.
Using our model, workflow, and explanations in \cite{Schons2017}, third parties are able to create additional scenarios.
In addition, our implementation is open to extensions such as more detailed activities, vehicle support, etc.

%% file: evaluation.tex

\begin{figure*}
	\newcommand{\heightpdf}{165pt}
	\newcommand{\hightpng}{161.5pt}
	\newcommand{\heightraise}{1.75pt}
	\includegraphics[height=\heightpdf]{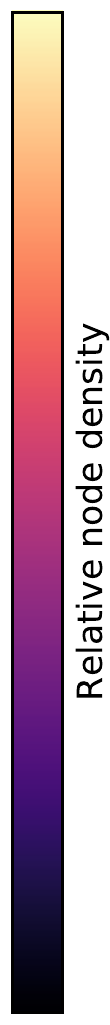}
	\hfill
	\subfloat[\gls{RWP}]{\raisebox{1.5pt}{
		\includegraphics[height=\hightpng]{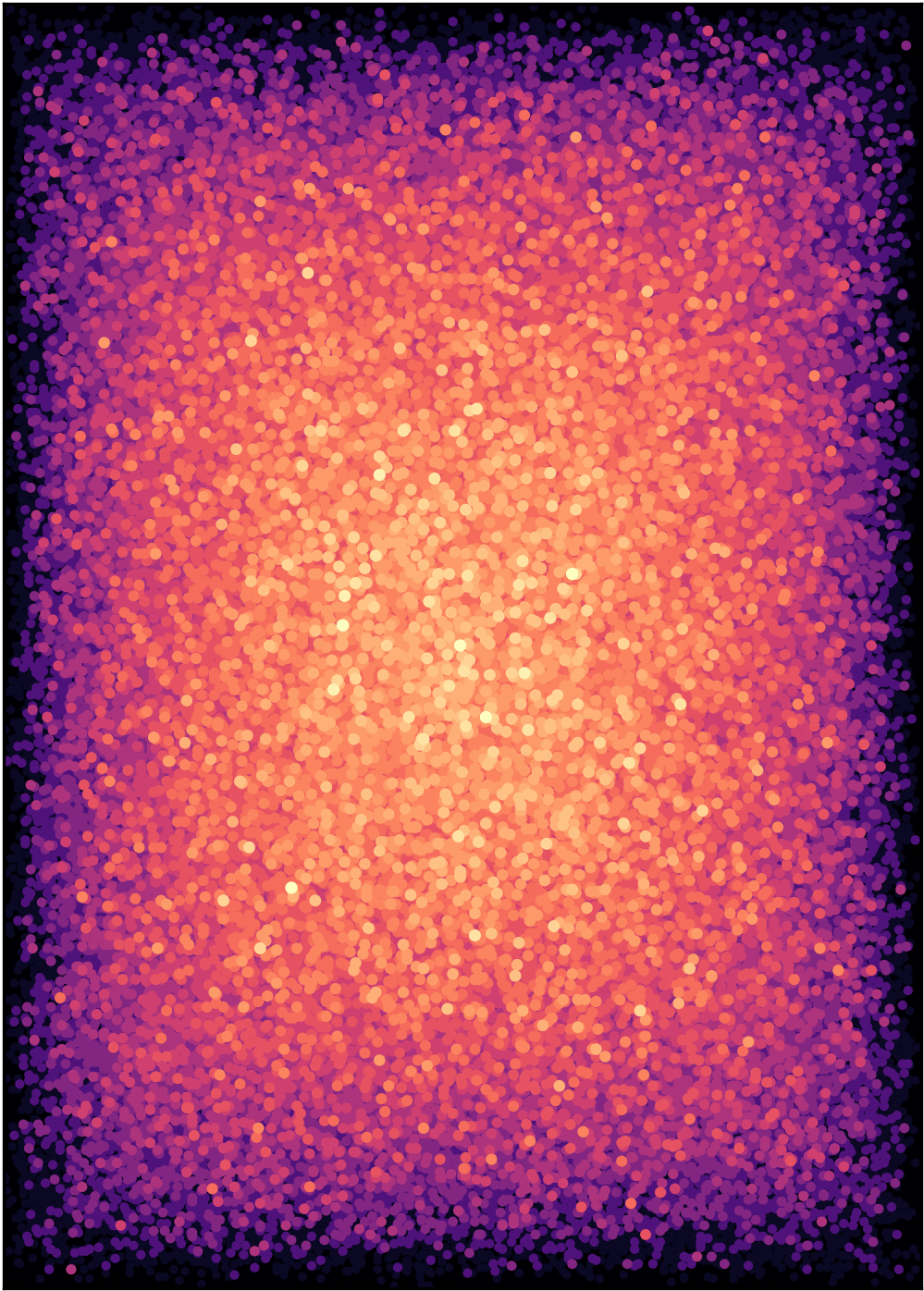}
		\label{fig:node-density-rwp}
	}}
	\hfill
	\subfloat[Map-based RWP]{\raisebox{1.5pt}{
		\includegraphics[height=\hightpng]{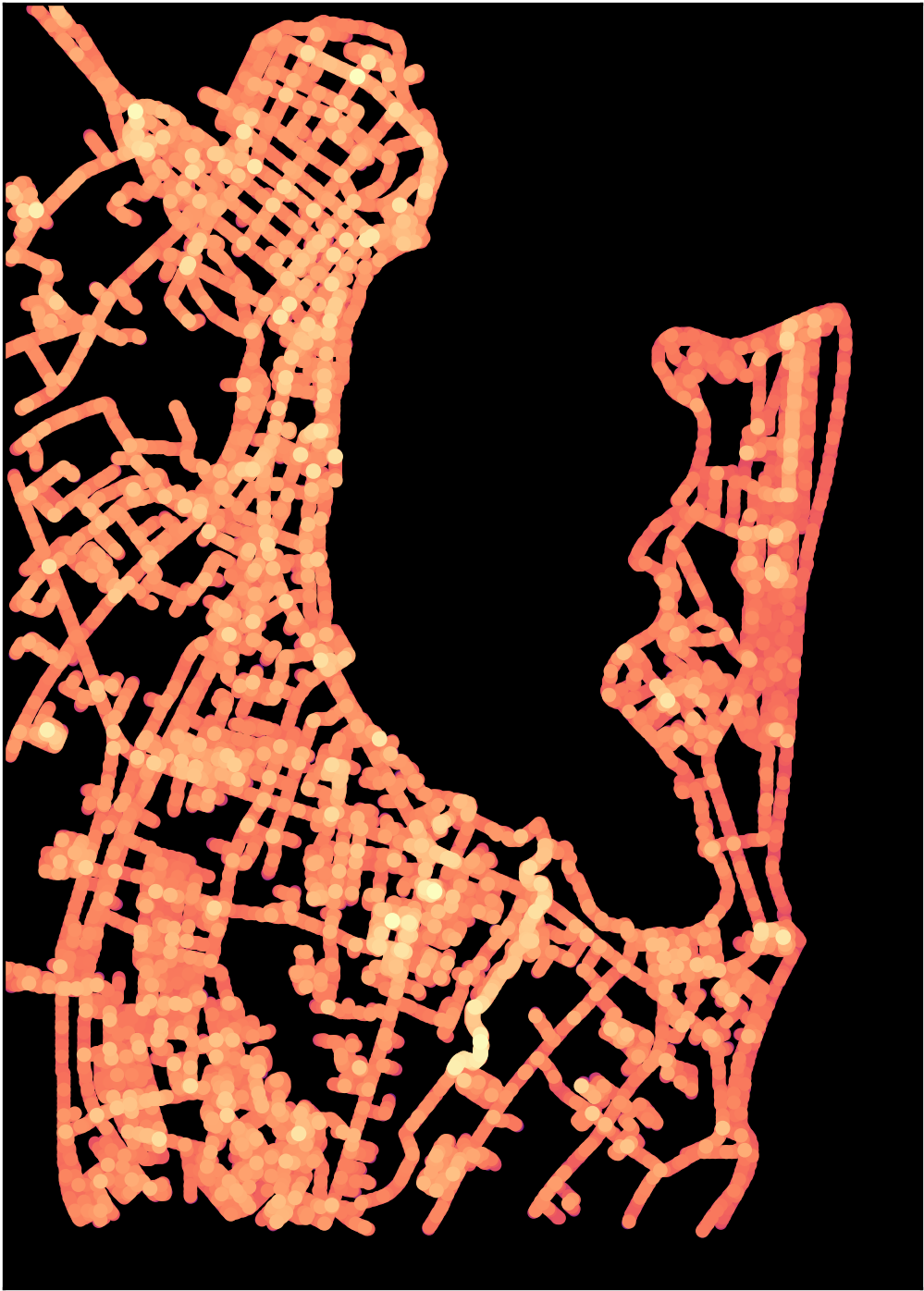}
		\label{fig:node-density-mapbased}
	}}
	\hfill
	\subfloat[Natural Disaster]{\raisebox{1.5pt}{
		\includegraphics[height=\hightpng]{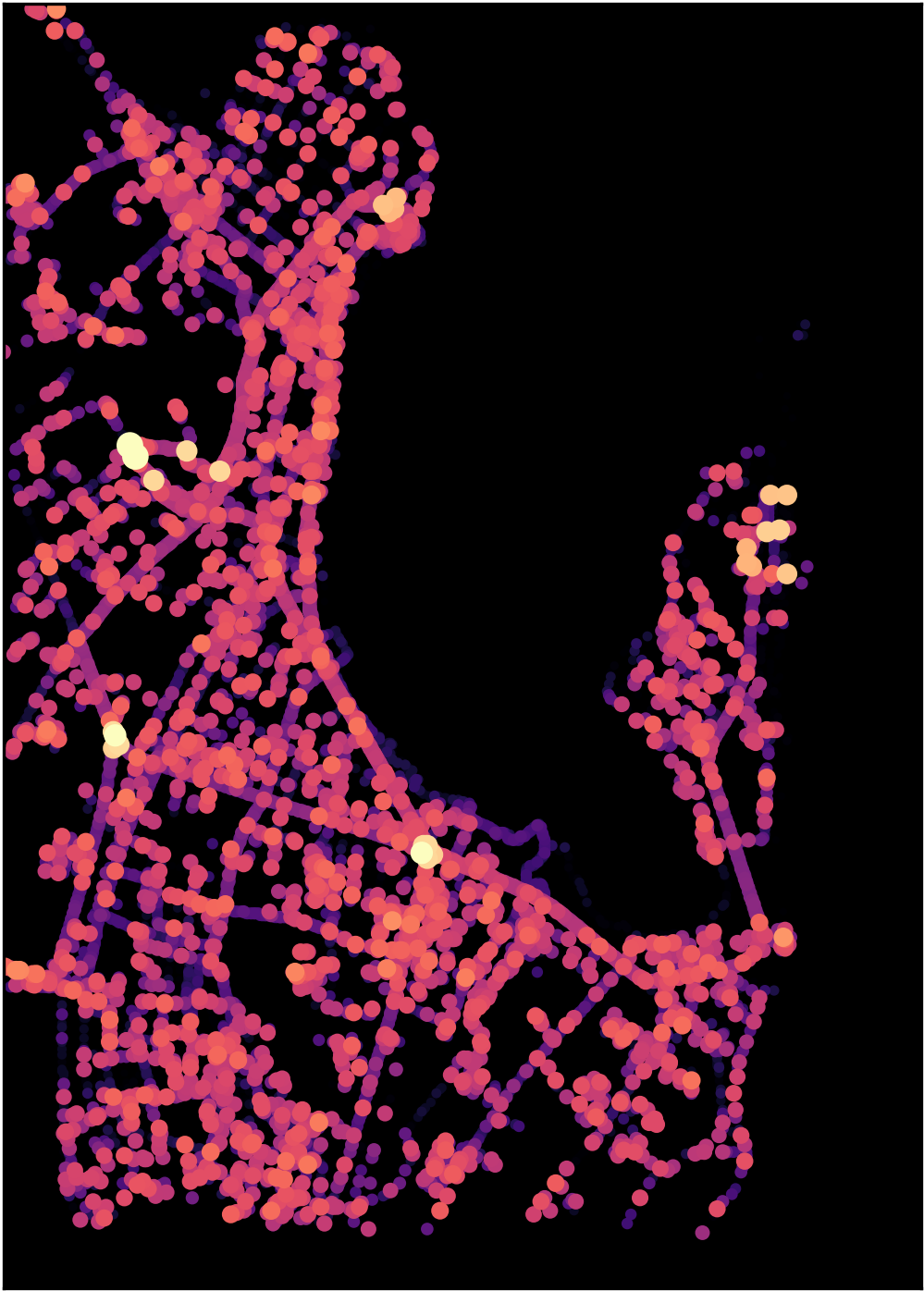}
		\label{fig:node-density-natdis}
	}}
	\hfill
	\subfloat[Reference map (same as \cref{fig:tacloban-map})]{%
		\includegraphics[height=\heightpdf]{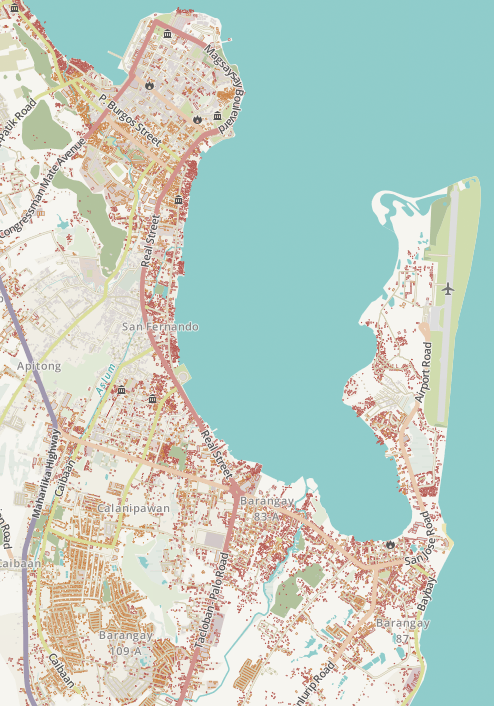}
		\label{fig:node-density-reference}
	}
	\caption{Spatial node distribution in different mobility models averaged over 7 days. Node counts are sampled from a grid of 10 $\times$ 10 m squares. Circle sizes and colors scale logarithmically with node count to highlight hot spots.}
	\label{fig:node-density-heatmap}
\end{figure*}

\section{Evaluation}
\label{sec:evaluation}

In this section, we want to
\begin{inparaenum}[($i$)]
	\item visually validate our mobility model (\emph{ND}),
	\item identify qualitative differences between \emph{ND} and other contemporary models,
	\item assess the impact that different mobility models have \gls{DTN} performance, and finally
	\item give actionable advice to \gls{DTN} protocol designers.
\end{inparaenum}
To this end, we compare \emph{ND} with two other widely used models:
the Random Waypoint Mobility model (\emph{\gls{RWP}}) and a map-based \gls{RWP} model (\emph{Map}) where waypoint selection is still random, but node movement is confined to a street grid.
We selected the \emph{epidemic} \gls{DTN} routing protocol for the network performance analysis as its simple design facilitates reasoning about the results.
For all experiments, we rely on the ONE simulator v1.6.0 \cite{keranen2009theone}. Each experiment is averaged over 10 independently seeded runs.
We summarize the most important simulation settings in \cref{tab:eval-simulator-settings} and refer to our source code \cite{NatdisCode} for all details. For reproducibility, we also publish our experiment data set~\cite{NatdisData}.

We note the limitation of our simulation setup with only 500 nodes. Even though our mobility model supports an arbitrary number of nodes, current network simulators (ONE, ns-2/3, etc.) do not scale to a large number of nodes, e.\,g., the population size in an urban area (order of 10$^6$). 
However, as we will show, we can demonstrate qualitative differences between the different mobility models already with a rather low number of nodes.
We also note that in our traffic model, sources and destinations as well as messages sizes are arbitrarily chosen. A realistic communication model would be incredibly useful but is orthogonal to our work.

\begin{table}
	\renewcommand{\arraystretch}{0.95} 
	\small
	\centering
	\caption{The ONE Simulation Settings}
	\label{tab:eval-simulator-settings}
	\begin{tabulary}{1\linewidth}{@{}rll@{}}
		\toprule
		
		\multirow{3}{*}{\emph{Scenario}}
		& Dimensions & 5000 $\times$ 7000\,m$^2$ (Tacloban) \\
		& Simulation time & 7 days \\
		& Nodes & 500 \\
		\midrule
		
		\multirow{2}{*}{\emph{Mobility}}
		& Model & \emph{RWP}, \emph{Map}, \emph{ND} \\
		& Speed (m/s) & 0.5--1.5 $\approx$ walking speed \\
		\midrule
		
		\multirow{2}{*}{\emph{Routing}}
		& Algorithm & \emph{Epidemic} \\
		& Buffer size & 20 MB \\
		\midrule
		
		\multirow{3}{*}{\emph{Taffic}}
		& Msg.\ interval & 8--12\,s \\
		& Msg.\ TTL & 6\,h \\
		& Msg.\ size & 50--100 KB\\
		\midrule
		
		\multirow{2}{*}{\emph{Comm.}}
		& PHY rate & 2 Mbit/s $\approx$ Bluetooth \\
		& Radio range & 10\,m \\
		
		\bottomrule
	\end{tabulary}
\end{table}

\subsection{Characterizing Mobility}

\begin{figure}[b]
	\centering
	\tikzset{style={font=\footnotesize}}
	\includegraphics[width=0.9\linewidth]{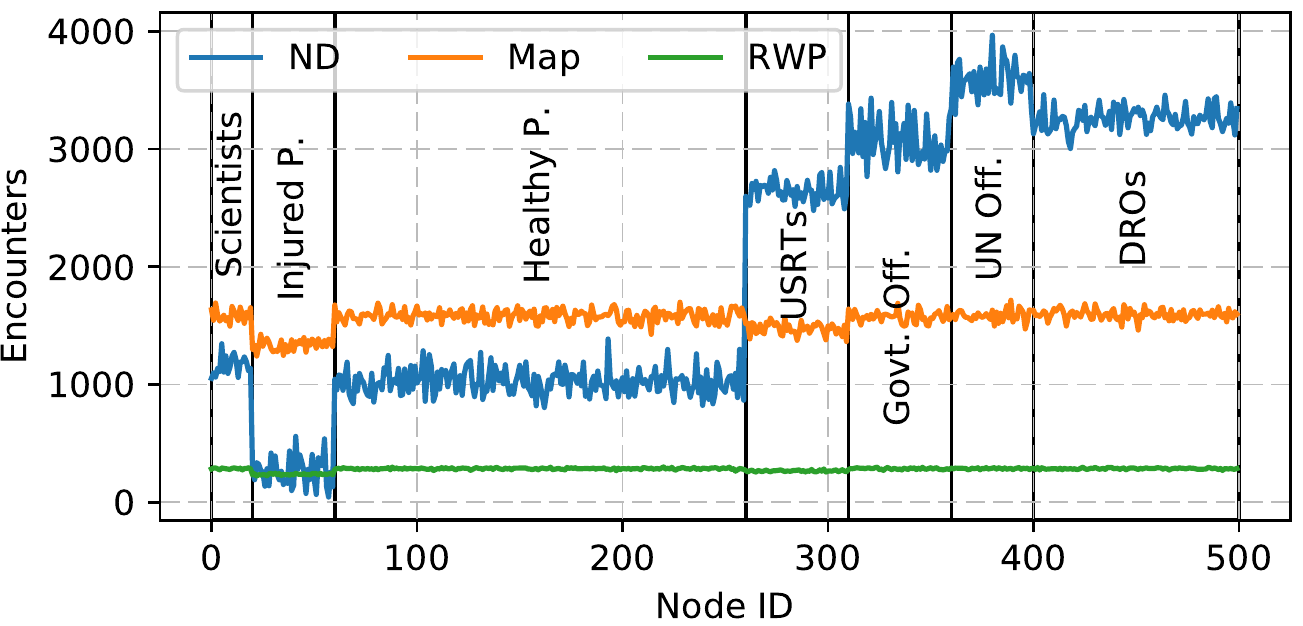}
	\caption{Number of encounters per node (1 week).}
	\label{fig:encounters-per-node}
\end{figure}

We are interested in the spatial node distribution and encounters that occur during a disaster since they both affect the applicability of a \gls{DTN}: node hot spots can function as communication hubs where messages are quickly exchanged, while nodes that have a lot of encounters can act as \enquote{data mules} and transport messages over larger distances.

\subsubsection{Spatial Node Distribution}
We visualize the spatial node distribution of the three mobility models using a scatter-plot heatmap in \cref{fig:node-density-heatmap}.
In \cref{fig:node-density-rwp}, we identify the typical non-uniform center-weighted distribution \cite{Bettstetter2003RWPdistribution} of the \emph{\gls{RWP}} model.
From a practical perspective, this means that nodes are moving across inaccessible areas, for example, a bay.
In contrast, \emph{Map} and \emph{ND} (\cref{fig:node-density-mapbased,fig:node-density-natdis}) essentially ``redraw'' the underlying street grid. Here, the nodes' movements are confined to streets and paths and are thus no longer moving across water.
However, node distribution in \emph{Map} across streets appears generally uniform. There are only minor hot spots at street intersections which are expected since movement trajectories cross there. In general, nodes are located with similar probabilities at any point in the map.
\Cref{fig:node-density-natdis} shows that \emph{ND} exhibits characteristic hot spots that can be mapped to certain \glspl{POI} in the street map (\cref{fig:node-density-reference}) where many nodes stay for a longer period of time.
Most prevalent are the locations of the OSOCC and the base camp as they are frequented by \glspl{DRT} and officials. In addition, we can also make out other hot spots at the city hall and the food and water distribution points.
In the simulation, the airport is the location for inactive nodes, that is, those nodes that have not yet arrived and those that have already left. The hot spots around the airport can therefore be considered an artifact as the simulator does not support removing nodes from a running simulation.

\subsubsection{Encounters}
An \emph{encounter} is a transmission opportunity which occurs if two nodes move in each other's transmission range.
\glspl{DTN} performance highly depends on the number of encounters a node makes while moving around. For example:
if a node encounters the destination of any currently carried message, it can directly deliver it. The advantage of direct delivery is that it prevents the replication overhead to intermediate nodes in form of radio transmissions and storage consumption. So, in a scenario where communicating parties are generally physically close to one another, or at least meet regularly, a \gls{DTN} deployment could exclusively rely on direct deliveries.
Therefore, assessing the encounter characteristics of the underlying mobility model in a scenario is of utmost importance to understand which protocols are suitable for a natural disaster scenario.
%
In \cref{fig:encounters-per-node}, we observe that in \emph{ND}, the local population groups (healthy and injured) make significantly fewer contacts than the other groups. Especially the injured encounter very few other nodes.
On the other hand, \gls{DRO} teams as well as government and UN officials make significantly more contacts due to regular meetings at the OSOCC, in the town hall, and in the base camps.
In the \emph{\gls{RWP}} and \emph{Map} models, the number of encounters solely depends on the average velocity of the user role. For example, injured as well as heavily equipped \glspl{USRT} move slower than the other groups.
The generally low number of encounters of \emph{\gls{RWP}} can be explained by the low node density in combination with the low transmission range: as the nodes freely move around the large area, nodes only infrequently move into each others transmission range.

\subsection{Characterizing Network Performance}
\begin{figure*}
	\begin{minipage}[t]{0.33\textwidth}
		\centering
		\includegraphics[width=\linewidth]{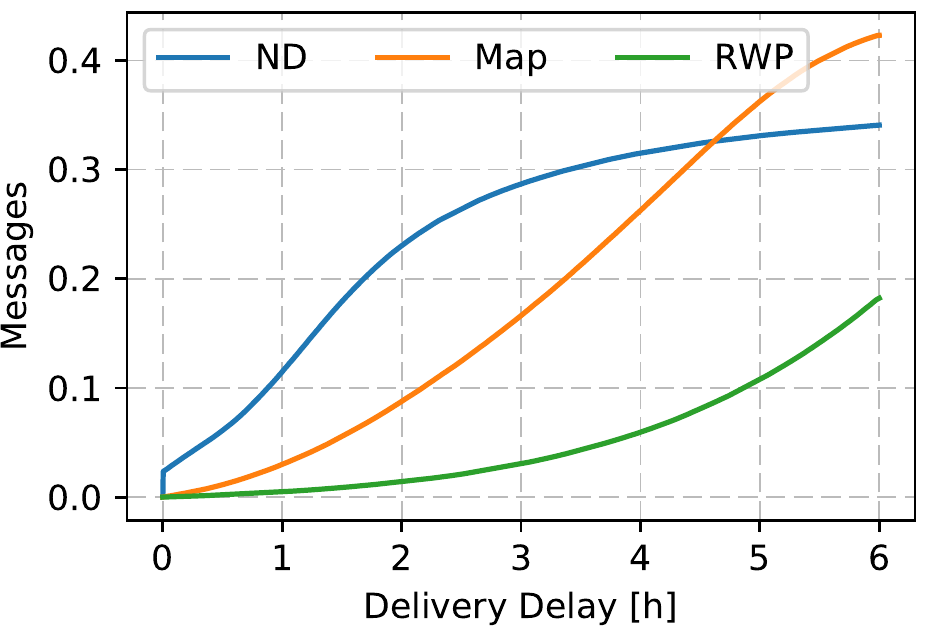}
		\caption{Message delivery success and delay as a cumulative distribution function.}
		\label{fig:delivery-cdf}
	\end{minipage}
	\hfill
	\begin{minipage}[t]{0.33\textwidth}
		\centering
		\includegraphics[width=\linewidth]{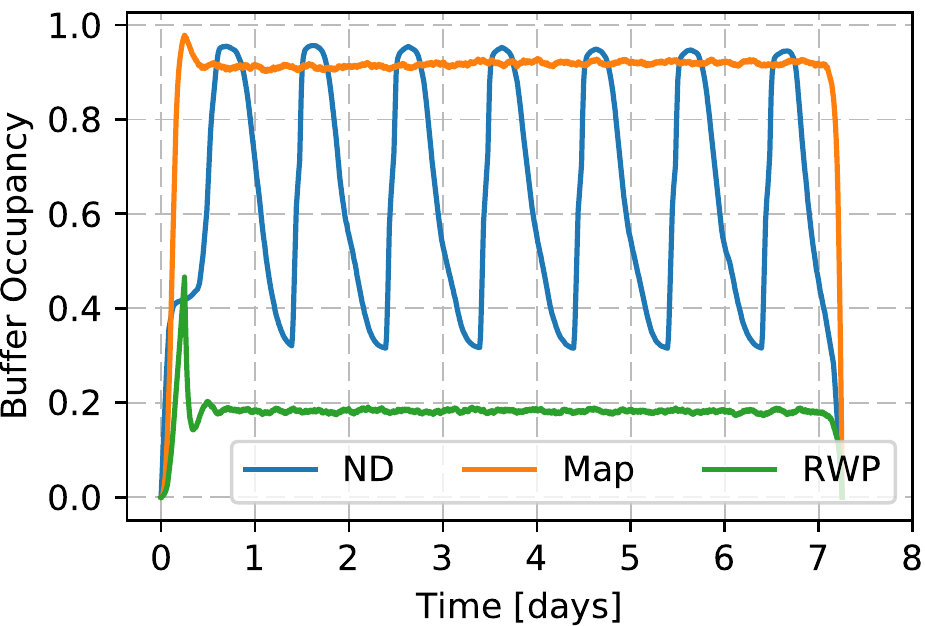}
		\caption{Buffer occupancy over time.}
		\label{fig:buffer-occupancy}
	\end{minipage}
	\hfill
	\begin{minipage}[t]{0.28\textwidth}
		\centering
		\includegraphics[width=0.95\linewidth]{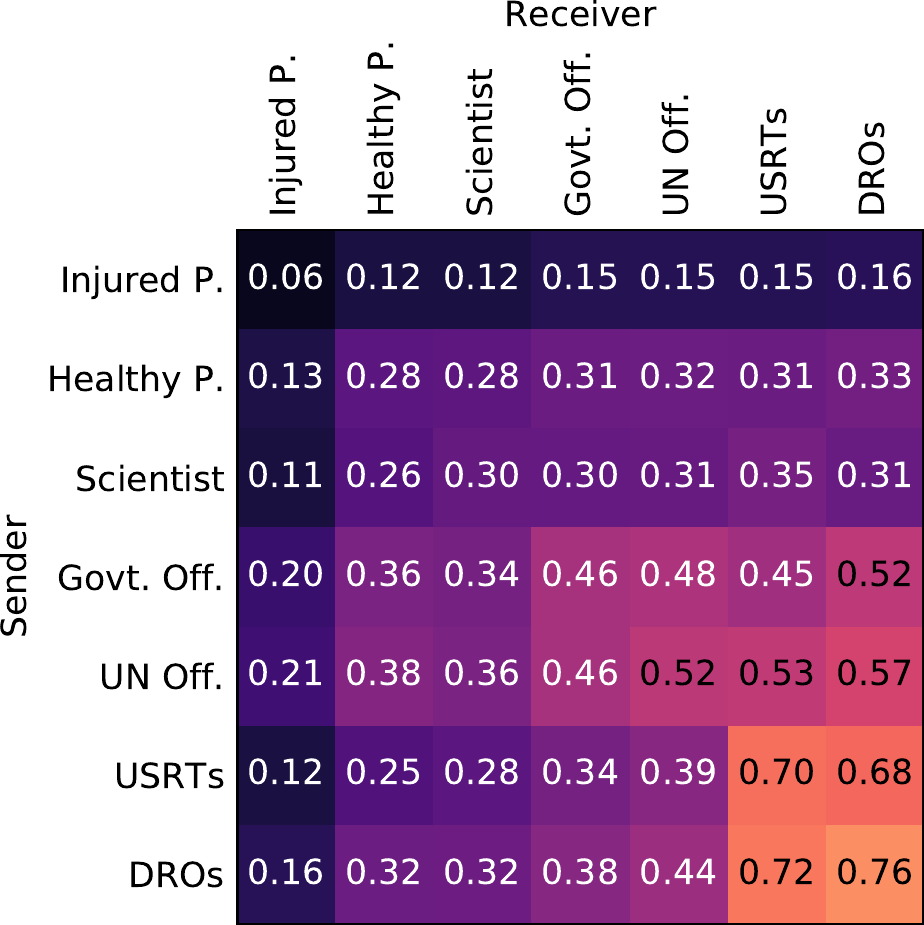}
		\caption{Delivery rates between roles (\emph{ND}).}
		\label{fig:delivery-matrix}
	\end{minipage}
\end{figure*}

We evaluate the impact of our mobility model on network performance.
For this, we measure
the delivery rate and delay,
the buffer occupancy,
as well as the delivery rate for the different roles.
In \cref{fig:delivery-cdf}, we observe that the delivery delay short in \emph{ND}, meaning that about 80\,\% of all delivered messages are delivered within three hours, which appears to be due to the regular meetings at certain \glspl{POI}.
However, we can see in \cref{fig:delivery-matrix} that delivery success is highly heterogeneous with respect to the user role in \emph{ND}. We observe that the most successful communication partners are the ones that make the most encounters, in particular, \glspl{DRO}, \glspl{USRT}, and government/UN officials.
These are the groups that can rely on message deliveries the most and are, thus, the main beneficiaries of the communication network.
On the other hand, less mobile nodes, especially the injured population, are poorly connected to the rest. Without special treatment (e.\,g., via prioritization), these users will be insufficiently served by the network.
In \emph{\gls{RWP}} and \emph{Map}, performance is almost uniform across the different group (not shown here).
The overall delivery success of \emph{ND} is lower than for \emph{Map} which is likely due to the fact that at night, when mobility is low, only few messages are delivered.
This is supported by \cref{fig:buffer-occupancy} which shows that buffer occupancy is depending on the time of day. During the night, undelivered messages expire and nodes remove them from their buffers. Starting around noon, buffers refill with new messages as nodes travel across the map and collect messages from others, leading to more message deliveries during the day.
Note that the delivery success over time is not shown here for space reasons, but follows the same cyclic behavior as the buffer occupancy.

%% file: conclusion.tex

\section{Conclusion}
\label{sec:conclusion}

\glspl{DTN} could greatly facilitate disaster response communications as they allow message-based communication even in the absence of supporting infrastructure.
However, so far, the communication community has yet to prove the practicality of \glspl{DTN} for such scenarios.
In this paper, we have shown the feasibility of reverse engineering human mobility solely by written or oral reports and have extract general as well as scenario-specific features and patterns.
Based on two exemplary scenario, we were able to assess the performance of \glspl{DTN} in large-scale natural disasters as a backup for infrastructure-based communication, and develop a tool for future works in this area.
In particular, we observed that a \gls{DTN} deployment could be most useful for disaster relief workers due to regular meetings, while the population benefits to a lesser degree.
Based on these observations, we give advice to \gls{DTN} designers such as preferring  frequently traveling nodes as relays and exploiting hot spots by deploying fixed \gls{DTN} nodes to serve as information exchange hubs.
In the future, we could increase the level of detail in our mobility model, for example, by including vehicles.
Our network performance evaluation could be extended, which would require to increase the scalability of current simulators to support simulations with significantly more nodes.
Finally, a realistic communication model of all users in a disaster scenario would complement our mobility model since communication patterns and user mobility are the two main factors influencing \gls{DTN} performance.